\documentclass[aps,prd,showpacs,amsmath,amssymb]{revtex4}
\usepackage{epsfig}
\begin{document}

\title{Zero temperature lattice Weinberg - Salam model for the values of
the cutoff  $\Lambda \sim 1$ TeV}

\author{M.A.Zubkov}
 \email{zubkov@itep.ru}
\affiliation{ITEP, B.Cheremushkinskaya 25, Moscow, 117259, Russia}

\today


\begin{abstract}
The lattice Weinberg - Salam model at zero temperature is investigated
numerically. We consider the model for the following values of the coupling
constants: the Weinberg angle $\theta_W \sim 30^o$, the fine structure constant
$\alpha \sim \frac{1}{150}$, the Higgs mass $M_H \sim 150$ GeV. We find that
the fluctuational region begins at the values of the cutoff $\Lambda$ above
about $0.8$ TeV. In this region the average distance between Nambu monopoles is
close to their sizes.  At $\Lambda > 1.1$ TeV the  Nambu monopole currents
percolate. Within the fluctuational region the considered definitions of the
scalar field condensate give values that differ from the expected one $2
M_Z/g_z$.  We consider the given results as an indication that the
nonperturbative effects may be present in the Weinberg - Salam model at the
large values of the cutoff. Our numerical results were obtained on the lattices
of sizes up to $16^3\times 32$.
\end{abstract}


\maketitle


\section{Introduction}

Investigation of the phase transitions often requires application of
nonperturbative methods. In particular, the nonperturbative phenomena are
important for the description of the finite temperature Electroweak phase
transition \cite{1,2,3,4,5,6,7,8,9,10,11,review,alg,EW_T}. At the same time,
the phase diagram of the lattice Weinberg-Salam model at zero temperature also
contains the phase transition \cite{Montvayold,13,14,Montvay}. The phase
transition surface separates the Higgs phase from the symmetric phase. On both
sides of the phase diagram it is necessary to find the way to approach
continuum physics within the lattice model. It is expected, that the continuum
physics arises in some vicinity of this transition (on different sides of the
transition different continuum models appear). Strictly speaking, this pattern
is self - consistent only if the transition is of the second order. In this
case close to the phase transition lattice spacing tends to zero and a kind of
a continuum field theory appears. Indeed, zeroth order of the perturbation
theory predicts the second order phase transition in the Weinberg - Salam
model. However, already on the one - loop level the Coleman - Weinberg
effective potential predicts the first order phase transition. This discrepancy
points out to numerical lattice methods as to the judge. There may take place
the 1-st order phase transition or the 2-nd order phase transition. Also
another possibility appears: the transition may appear to be a crossover.

That's why we expect nonperturbative effects to appear in the lattice Weinberg
- Salam model close to the transition between the two above mentioned phases.
Or, in the other words, we expect nonperturbative effects to appear above some
energy scale, because the increase of the energy scale corresponds to the
decrease of the lattice spacing and, therefore, is achieved when the phase
transition is approached. Basing on trivial dimensional analysis we may expect
that the mentioned scale can be compared to the Electroweak scale $\sim 250$
GeV. In fact, some indications were recently found that this scale might be
around $1$ TeV (see, for example, \cite{VZ2008, VZ2009L, BVZ2008, Z2009,
Z2010Q, Z2010, PZ2011}). Namely, in the Electroweak theory there exist objects
that are not described by the first orders of the perturbation theory: Nambu
monopoles and the Z - strings \cite{Nambu}. It has been found that there exists
the vicinity of the phase transition \cite{Z2010}, where the average distance
between the Nambu monopoles is compared to their sizes. This region was called
in \cite{Z2009} the fluctuational region (FR). Nambu monopoles may be
considered as embryos of the unphysical phase within the physical one.
Therefore, it is natural to suppose that within the FR both phases are mixed
and neither the perturbation expansion around vacuum with zero scalar field nor
the perturbation expansion around vacuum with nonzero scalar field cannot give
the correct description of the situation. Besides, in \cite{PZ2011} it was
shown that there exist different ways to define scalar field condensate that
give identical values out of the FR, and different values within the FR. On the
boundary of the FR the lattice spacing  $a$ remains finite and practically does
not depend on the lattice size (for the considered lattices). Actually, the
value of the ultraviolet cutoff $\frac{\pi}{a}$ on this boundary for the values
of the Higgs boson mass $100, 150, 300$ GeV is around $1$ TeV. Thus the
hypothesis was suggested that above the energy scale $1$ TeV in the Weinberg -
Salam model nonperturbative effects may become important. It is worth
mentioning that these effects, most likely, are related to the expansion in
powers of $\lambda$ while the first orders of the perturbation theory for the
expansion in powers of $\alpha$ are expected to stay at work. In particular, no
discrepancy was found between the renormalized fine structure constant and its
one - loop estimate \cite{PZ2011}.

In the present paper we extend the research of \cite{PZ2011} to larger lattices
(in \cite{PZ2011} the lattices of sizes $8^3\times 16$ were used; here we use
lattices $16^3\times 32$). In addition we investigate the properties of Nambu
monopoles and $Z$ - strings that were out of the scope of the mentioned above
papers. Namely, we consider their percolation properties that are related to
their possible condensation. In the finite temperature theory it was found that
the Electroweak transition is accompanied with the condensation of the Nambu
monopoles and the condensation of the Z - strings
\cite{Chernodub_Nambu,Gubarev, BVZ2006}. Here we find that in the zero -
temperature model this occurs as well. The "percolation transition" in the
Weinberg - Salam model at the values of couplings we consider is situated
within the FR. And there exists the subregion of the FR, were both Z - strings
and Nambu monopoles are condensed, while at least one of the considered
definitions of the scalar field condensate still gives nonzero value.

 The paper is organized as
follows. In Section 2 we consider the definition of the lattice regularized
Weinberg - Salam model and describe the details of the simulation. In Section 3
we discuss the phase diagram of the lattice model and the lines of constant
physics. In Section 4 we describe how lattice spacing was calculated in our
study. In Section 5 we calculate the renormalized fine structure constant. In
Section 6 we investigate three different scalar field condensates. In Section 7
we calculate Z - string and Nambu monopole percolation probabilities. In
Section 8 we discuss the obtained numerical results. Throughout the paper the
notations of differential forms on the lattice are used (for their definition
see, for example, \cite{forms}).

\section{The lattice regularized Weinberg - Salam model}

We consider the model without fermions. Its partition function has the form:

\begin{equation}
Z = \int D H D\Gamma exp(-A(\Gamma,H))
\end{equation}

Here $A(\Gamma,H)$ is the action for the scalar doublet $H$ and the gauge field
$\Gamma = U\otimes e^{i\theta} \in SU(2)\otimes U(1)$:
\begin{eqnarray}
 A(\Gamma,H) & = & \beta \!\! \sum_{\rm plaquettes}\!\!
 ((1-\mbox{${\small \frac{1}{2}}$} \, {\rm Tr}\, U_p )
 + \frac{1}{{\rm tg}^2 \theta_W} (1-\cos \theta_p))+\nonumber\\
 && + \frac{\gamma}{2} \sum_{xy} |H_x - U_{xy} e^{i\theta_{xy}}H_y|^2 + \sum_x (|H_x|^2 (1 - 2\lambda - 4\gamma) +
 \lambda |H_x|^4), \label{S}
\end{eqnarray}

On the tree level we have:
\begin{eqnarray}
v &=& \sqrt{2\frac{\gamma - \gamma_c}{\lambda}}\nonumber\\
m_H &=& v\sqrt{\frac{8\lambda}{\gamma}}\nonumber\\
m_Z &=& v\sqrt{\frac{\gamma}{\beta \, {\rm cos}^2 \theta_W}}\nonumber\\
\gamma_c &=& \frac{1 - 2\lambda}{4}\label{tree}
\end{eqnarray}
Here we have introduced: vacuum expectation value $v$ of $|H_x|$, the lattice
Higgs boson mass $m_H = M_H a$, the lattice $Z$ - boson mass $m_Z = M_Z a$, and
the critical value $\gamma_c$.

After fixing Unitary gauge
\begin{equation}
H = \left(\begin{array}{c}h\\0\end{array}\right), h \in R,\label{unitary}
\end{equation}
where $H$ is the scalar doublet, the $Z_2$ gauge ambiguity remains: $h_x
\rightarrow (-1)^{n_x} h_x, \, Z \rightarrow [Z + \pi d n]{\rm mod}\, 2 \pi$.
Here  the $Z$ - boson field is defined as
\begin{equation}
Z = -{\rm Arg}\, [U_{11}e^{i\theta}] \label{Z0_}
\end{equation}

 The tree level approximation gives for the infrared effective constraint
potential \cite{PZ2011} after any $Z_2$ gauge is fixed:
\begin{equation}
V^{i-r}(\phi) =  N_4 \lambda (\phi^2 - v^2 )^2 \label{IRPOT_}
\end{equation}
Here $N_4$ is the lattice volume.

In numerical simulations we use Metropolis algorithm. The model is simulated in
Unitary gauge with the signs of $h$ unfixed. After each $150$  Metropolis
sweeps the Z (or DZ) - version of Unitary gauge is fixed (for the definition of
these gauges see the next sections). As a starting point of our simulations on
the lattice $16^3\times 32$ we use configurations obtained on the lattice
$8^3\times 16$ during the preparation of \cite{PZ2011}. $16$ identical
configurations are merged together forming the starting $16^3\times 32$
configuration. Then, about $ 60 000$ Metropolis sweeps are made before the
measurement of observables begins (this has required about $600$ ours CPU
time). During this preliminary run the 16 mentioned above parts of the lattice
become decorrelated which signalizes that the thermalization is achieved.

\section{Phase diagram and lines of constant physics}

\begin{figure}
\begin{center}
 \epsfig{figure=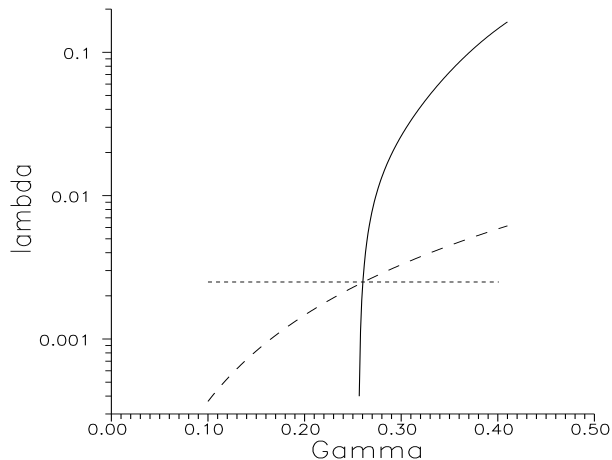,height=60mm,width=80mm,angle=0}
\caption{\label{phase} The phase diagram of the model in the
 $(\gamma, \lambda)$-plane at $\beta = 12$. The dashed line is the tree - level
  estimate for the line of constant physics ($\frac{\lambda}{\gamma^2} = \frac{1}{8 \beta} \frac{M^2_H}{M^2_W} = {\rm
const}$) correspondent to bare $M^0_H = 150$ Gev. The continuous line
  is the line of the phase transition between the physical Higgs phase and the unphysical symmetric phase (statistical errors for
  the values of $\gamma$ at each $\lambda$ on this line are about $0.005$).  The dotted line is the line $\lambda = 0.0025$. Along this line
  the physical quantities are calculated in the present research. }
\end{center}
\end{figure}

The lattice model defined by Eq. (\ref{S}) has the four - dimensional ($\beta,
\gamma, \lambda, \theta_W$) phase diagram. On this phase diagram phase
transition surface is three - dimensional. The lines of constant physics on the
tree level are the lines ($\frac{\lambda}{\gamma^2} = \frac{1}{8 \beta}
\frac{M^2_H}{M^2_W} = {\rm const}$; $\beta = \frac{1}{4\pi \alpha}={\rm
const}$; $\theta_W = {\rm const}$). We suppose that in the small vicinity of
the transition  the deviation of the lines of constant physics from the tree
level estimate may be significant. However,  qualitatively their behavior is
the same. Namely, the cutoff is increased along the line of constant physics
when $\gamma$ is decreased and the maximal value of the cutoff is achieved at
the transition point. Nambu monopole density in lattice units is also increased
when the ultraviolet cutoff is increased.

 In our lattice study we fix bare $\theta_W = \pi/6, \beta = 12, \lambda = 0.0025$.
Therefore, strictly speaking we investigate the system along the line on the
phase diagram that differs from the line of constant physics. This is
illustrated by Fig. \ref{phase}, where the projection of the phase diagram to
the plane ($\beta = 12, \theta_W = \pi/6$) is drawn.  This diagram is obtained,
mainly, using the lattice $8^3\times 16$. Some regions ($\lambda =
0.009,0.0025, 0.001$), however, were checked using larger lattices (see, for
example, \cite{VZ2008,VZ2009L,Z2009,Z2010,Z2010Q}). According to our data there
is no dependence of the diagram on the lattice size. The physical Higgs phase
is situated right to the solid transition line. The position of this line was
determined using various methods \cite{VZ2008,VZ2009L,Z2009,Z2010,Z2010Q}.
However, the uncertainty is still present in the final determination of the
phase transition points. This uncertainty on this Figure is within the error
bars of $\gamma = \gamma_c \pm 0.005$. The details of this uncertainty at
$\lambda = 0.0025$  are discussed in the present paper (see below, this
Section). The dashed line represents the tree - level estimate for the line of
constant physics. The dotted line is the line $\lambda = 0.0025$. Along this
line the physical quantities are calculated that are reported in the present
paper. One can see, that already on the tree level this straight line deviates
from the line of constant physics. However, Figure \ref{phase} demonstrates
also that within the interval $\gamma \in [0.255, 0.27]$ (where the physical
quantities are measured in the present research) the deviation of the tree
level estimate for the line of constant physics from the straight line $\lambda
= 0.0025$ is not crucial. In fact, on the tree level along this straight line
the fine structure constant does not vary. The renormalized fine structure
constant is also almost not changed (see discussion below in Section
\ref{sectfine}). As for the Higgs boson mass, its value on the tree level
varies between $154$ GeV at $\gamma = 0.255$ and $145$ GeV at $\gamma = 0.27$.
The variation of the renormalized Higgs mass along the line $\lambda = 0.0025$
is discussed in Section \ref{sectz}. Our data (with large statistical errors,
though) also demonstrate that the Higgs mass does not deviate significantly
from bare value $\sim 150 $ GeV at $\lambda = 0.0025, \gamma \in [0.2585,
0.27]$. It is worth mentioning that we did not investigate the renormalized W -
boson mass in the present research. Therefore we do not represent here any data
on the renormalized Weinberg angle. However, we also expect that it does not
vary sufficiently at the considered values of $\gamma$.

In the present paper we deal with the line $\lambda = 0.0025, \theta = \pi/6,
\beta = 12$ for $\gamma \in [0.2585, 0.27]$ as with an approximation of the
line of constant physics because along this line the main physical quantities
(Higgs mass, fine structure constant) do not vary essentially. The lowest value
$\gamma = 0.2585$ in the mentioned above interval is chosen because we expect
that for the description of the model at $\gamma < 0.2585$ larger lattice sizes
are necessary. This is related to the fact that at $\gamma < 0.2585$ the value
of the cutoff is larger than $1.5$ TeV and increases very fast (see discussion
below, in Section \ref{sectz}, Eq. (\ref{cutoffexpr})). At the same time
already for the values of the cutoff $\Lambda \sim 10$ TeV we need lattices of
linear size $>> 10$ (see discussion in Section \ref{disc}). We expect that the
line of constant physics at $\Lambda
>> 1$ TeV  deviates
essentially from the straight line investigated in the present paper. Not only
the Higgs mass may deviate from its bare value but also the renormalized fine
structure constant and, probably, the renormalized Weinberg angle. However, for
the investigation of such large values of $\Lambda$ extremely large lattices
are needed and such a research is out of the scope of the present paper.

\begin{figure}
\begin{center}
 \epsfig{figure=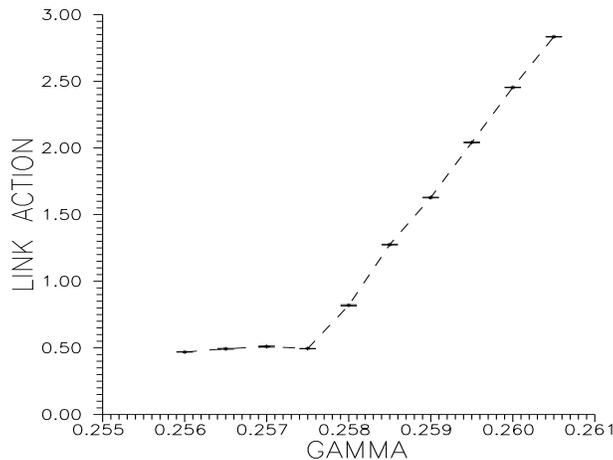,height=60mm,width=80mm,angle=0}
\caption{\label{LAQ} The link part of the action as a function of $\gamma$ at
$\lambda =0.0025$ , $\beta = 12$ on the lattice $16^3\times 32$.  }
\end{center}
\end{figure}

Below in the present paper we always deal with the investigation of the lattice
model at fixed $\theta_W = \pi/6, \beta = 12, \lambda = 0.0025$. In Fig.
\ref{LAQ} the data of the the link part of the action
$\frac{1}{4N_4}\sum_{xy}H_x^+U_{xy} e^{i\theta_{xy}}H_y$ are represented. The
dependence of the link part of the action on $\gamma$ indicates that the phase
transition (or, a crossover) can be localized at $\gamma \sim \gamma_c^{\prime}
= 0.25775 \pm 0.00025$. (See also \cite{PZ2011}, where the same conclusion was
made basing on the data obtained on the lattice $8^3\times 16$.) As in
\cite{PZ2011} we exclude the first order phase transition because we do not
observe any sign of a two - state signal. In the next sections we shall
demonstrate that the infrared UZ potential for the scalar field also points out
to $\gamma_c^{\prime}$ as to the transition point.

\begin{figure}
\begin{center}
\epsfig{figure=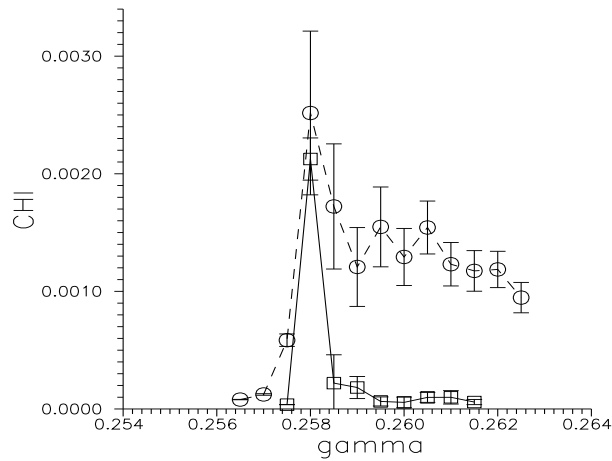,height=60mm,width=80mm,angle=0} \caption{\label{chi}
Susceptibility $\chi$ (in lattice units) as a function of $\gamma$ at $\beta =
12$, $\lambda = 0.0025$. Circles correspond to lattice $8^3\times 16$ while
squares correspond to lattice $16^3\times 32$.  }
\end{center}
\end{figure}

In addition in Fig. \ref{chi} we represent the fluctuation of the zero momentum
scalar field $\chi = [\delta H_{p=0}]^2 = <[H_{p=0}]^2> - <H_{p=0}>^2$ in
lattice units as a function of $\gamma$. Here $H_{p=0} = \frac{1}{N_4} \sum_x
|H_x|$. From this plot we obtain the critical value $\gamma_c^{\prime\prime} =
0.258\pm 0.0005$. Moreover, the given plot indicates that we deal with the
second order phase transition localized at $\gamma_c^{\prime\prime}$. Indeed,
the values of the fluctuation $\delta H_{p=0}$ on the lattice $8^3\times 16$
are about $\sqrt{2^4} = 4$ times larger than on the lattice $16^3\times 32$ for
$\gamma > \gamma_c^{\prime\prime}$. This means that for these values of
$\gamma$ the physical size of the lattice $8^3\times 16$ is twice smaller than
that of the lattice $16^3\times 32$, as it should when the correlation length
is smaller than the lattice size. However, at $\gamma_c^{\prime\prime}$ the
fluctuations calculated using both lattices almost coincide with each other.
This means that at $\gamma = \gamma_c^{\prime\prime}$ the physical sizes of
both lattices are the same, that may happen only if the correlation length is
much larger than the lattice size. It is worth mentioning, that the difference
between $\gamma_c^{\prime}$ and $\gamma_c^{\prime\prime}$ is $0.00025$ that is
within the statistical errors of both quantities. We feel this appropriate to
refer to $\gamma_c^{\prime}$ as to the possible phase transition point.

It is worth mentioning that Fig. \ref{chi} allows to estimate the fluctuation
of the scalar field within the fluctuational region. On its boundary, at
$\gamma = 0.2625$ (see below) we have $\chi \sim 0.001$ on the lattice
$8^3\times 16$. Also we know that at this value of $\gamma$ the correlation
length for the scalar field is about two lattice spacings. Therefore the given
lattice contains $4^3\times 8 = 512$ cubes of the linear size equal to the
correlation length. This means that the fluctuation of the scalar field within
such a cube is $\delta |H| \sim \sqrt{512 * \chi} \sim 0.7$ that is to be
compared with the average value  $\langle |H|\rangle \sim 2$. The formal
requirement for the perturbation theory to be applied is $2 \sim \langle
|H|\rangle >> \delta |H| \sim 0.7$. It seems that this inequality is not
satisfied. For $\gamma < 0.2625$ the situation is even worse. For example, at
$\gamma = 0.26$ we have $\langle |H|\rangle \sim 2$ while $\delta |H| \sim
0.9$.

\section{Z - boson mass, lattice spacing, and Higgs boson mass}
\label{sectz}

\begin{figure}
\begin{center}
 \epsfig{figure=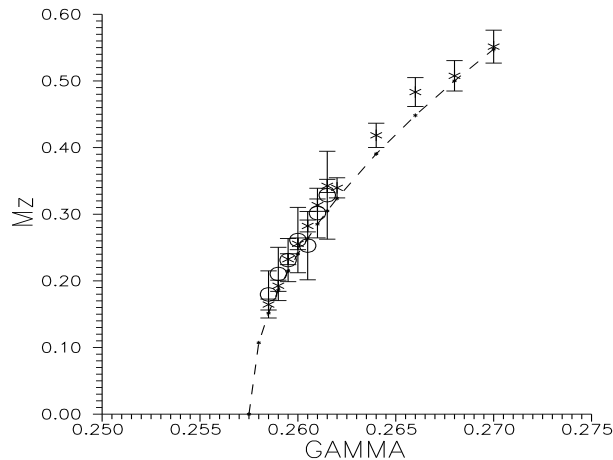,height=60mm,width=80mm,angle=0}
\caption{\label{uz} Z - boson mass in lattice units as a function of $\gamma$
at $\lambda =0.0025$ , $\beta = 12$. Crosses correspond to lattice $8^3\times
16$. Circles correspond to lattice $16^3\times 32$.  }
\end{center}
\end{figure}

For the calculation of the Z - boson mass we use the following definition of
the $Z$ boson field:
\begin{equation}
Z_{xy} = Z^{\mu}_{x} \;
 = - \,[{\rm Arg} (\Phi_x^+U_{xy} e^{i\theta_{xy}}\Phi_y) ]. \label{Z1_}
\end{equation}
Actually this definition of the Z - boson field coincides with the previous one
(\ref{Z0_}) taken in the version of Unitary gauge (\ref{unitary}) with
nonnegative $h$.

In order to evaluate the mass of the $Z$-boson we use the correlator:
\begin{equation}
\frac{1}{N^6} \sum_{\bar{x},\bar{y}} \langle \sum_{\mu} Z^{\mu}_{x} Z^{\mu}_{y}
\rangle \sim
  e^{-m_{Z}|x_0-y_0|}+ e^{-m_{Z}(L - |x_0-y_0|)}
\label{corZ}
\end{equation}
Here the summation $\sum_{\bar{x},\bar{y}}$ is over the three "space"
components of the four - vectors $x$ and $y$ while $x_0, y_0$ denote their
"time" components. $N$ is the lattice length in "space" direction. $L$ is the
lattice length in the "time" direction. The full lattice 4 - volume is  $N_4 =
N^3 \times L$.

In order to evaluate the Higgs boson mass we use the correlator:
\begin{equation}
\frac{1}{N^6} \sum_{\bar{x},\bar{y}} \{\langle  |H_{x}| |H_{y}| \rangle -
\langle  |H_{x}| \rangle^2\} \sim
  e^{-m_{H}|x_0-y_0|}+ e^{-m_{H}(L - |x_0-y_0|)}
\label{corZ}
\end{equation}

We can roughly evaluate the dependence of the lattice Z - boson mass on the
lattice size as follows. In finite temperature theory gauge boson thermal
masses appear of the order of $m_g = M_g a \sim g T a \sim \frac{g}{N_T} $,
where $T$ is the temperature while $N_T$ is the lattice size in imaginary time
direction. Analogy to the finite temperature theory allows us to evaluate the
finite volume contribution to the Z - boson mass as $\Delta m_Z \sim
\frac{g_Z}{N}$, where $N$ is the linear lattice size. For $\alpha \sim 1/150, N
\sim 8$ we have $\Delta m_Z \sim 0.08$ while at $N \sim 16$ we expect $\Delta
m_Z \sim 0.04$.

In Fig. \ref{uz} we represent our data on the Z - boson mass. Our numerical
results confirm the results of \cite{Z2010}. Nonzero values of $Z$ - boson mass
are obtained at $\gamma \ge 0.2585$. At the same time for $\gamma \le 0.258$ we
observe large statistical errors for the $ZZ$ correlator. Therefore, in this
region of the phase diagram the $Z$ - boson mass cannot be calculated and we
suppose it vanishes somewhere between $\gamma = 0.25$ and $\gamma =
\gamma_c^{\prime}$.

Taking into account expression (\ref{tree}) for the Z - boson mass we use the
following fit ($\gamma_c$ is changed to $\tilde{\gamma_c^{\prime}} = 0.2575$)
to the data of Fig. \ref{uz} :
\begin{equation}
m_Z = \sqrt{\frac{2\gamma(\gamma - \tilde{\gamma_c^{\prime}})}{\beta \lambda \,
{\rm cos}^2 \theta_W}}\label{fitmz}
\end{equation}
This fit is represented on the plot by the dashed line. It is worth mentioning
that  $\tilde{\gamma_c^{\prime}}$ is within the error bars of
$\gamma_c^{\prime}$. We find that  (\ref{fitmz}) with this value substituted
instead of $\gamma_c$ approximates the data better than with $\gamma_c^{\prime}
= 0.25775$ or $\gamma_c^{\prime\prime} = 0.258$. This does not mean, that Fig.
\ref{uz} points out to $\tilde{\gamma_c^{\prime}} = 0.2575$ as to the
transition point instead of $\gamma_c^{\prime} = 0.25775$. Instead, this means
that there exist also other contributions to the dependence of the Z  - boson
mass on $\gamma$ in addition to the tree level estimate with the real critical
value of $\gamma$ substituted instead of its naive estimate.

Using the value of lattice $Z$ - boson mass $m_Z$ we can evaluate the
ultraviolet cutoff $\Lambda = \frac{\pi}{a}$ as a function of $\gamma$ via the
relation $m_Z = a \times 91$ GeV. Further we shall use fit (\ref{fitmz})  in
order to represent our results as a function of $\Lambda$. Namely, we use the
following representation for $\Lambda$:
\begin{equation}
\Lambda = \pi \sqrt{\frac{\beta \lambda \, {\rm cos}^2 \theta_W}{2\gamma(\gamma
- 0.2575)}}\, * \, 91\, {\rm GeV}  \label{cutoffexpr}
\end{equation}

\begin{figure}
\begin{center}
 \epsfig{figure=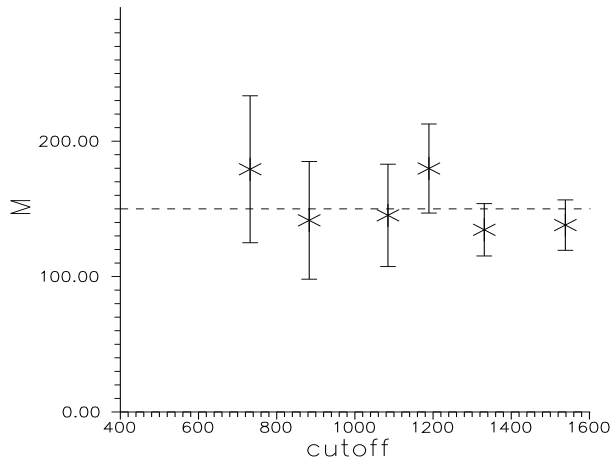,height=60mm,width=80mm,angle=0}
\caption{\label{fm} Higgs - boson mass in physical  units as a function of the
cutoff at $\lambda =0.0025$ , $\beta = 12$ on the lattice $8^3\times 16$.}
\end{center}
\end{figure}

 In
particular, in Fig. \ref{fm} we represent the Higgs boson mass calculated on
the lattice $8^3\times 16$ in physical units (in GeV) as a function of the
cutoff. Unfortunately, we do not have enough statistics to calculate this mass
on the larger lattices. We observe that the calculated values of the mass are
close to
 the expected value $\sim 150$ GeV. The deviation is
within the statistical errors.

Our estimate for the ultraviolet cutoff at $\gamma_c$ is $\sim 1$ TeV.
 The value of Z - boson
mass in lattice units at this point is about $\sim 0.2$. The above mentioned
estimate of the finite volume effect is $\sim 0.04$. Therefore we expect the
values of the ultraviolet cutoff reported here cannot differ from that of
obtained on an ideal infinite lattice by more than $20$ per cent. Thus we give
here the following estimate for the cutoff $\Lambda $ at $\gamma = \gamma_c$:
$\Lambda_c = 1 \pm 0.02$ TeV.

\section{Renormalized fine structure constant}
\label{sectfine}

In order to calculate the renormalized fine structure constant $\alpha_R =
e^2/4\pi$ (where $e$ is the electric charge) we use the correlator of Polyakov
lines for the right-handed external leptons. These lines are placed along the
selected direction (called below the imaginary time direction). The space -
like distance between the lines is denoted by $R$.
\begin{equation}
 {\cal C}(|\bar{x}-\bar{y}|)  =
 \langle {\rm Re} \,\Pi_{t} e^{2i\theta_{(\bar{x},t)(\bar{x},t+1)}}\,\Pi_{t} e^{-2i\theta_{(\bar{y},t)(\bar{y},t+1)}}\rangle.
\end{equation}
The potential is extracted from this correlator as follows
\begin{equation}
 {\cal V}(R) = -\frac{1}{L} { \rm log}\,  {\cal C}(R)
\end{equation}
Here $L$ is the size of the lattice in imaginary time direction.

Due to exchange by virtual photons and Z - bosons one would expect the
appearance of the Coulomb and Yukawa interactions:
\begin{eqnarray}
 {\cal V}(r) & = & -\alpha_R \, [{\cal U}_0(r)+\frac{1}{3}{\cal U}_{m_Z}(r)] +   const,\,
\nonumber\\
{\cal U}_m(r) & = & -\frac{ \pi}{N^3}\sum_{\bar{p}} \frac{e^{i p_3 r}}{{\rm
sin}^2 p_1/2 + {\rm sin}^2 p_2/2 + {\rm sin}^2
 p_3/2 +  {\rm sh}^2 m/2}
 \label{V2}
\end{eqnarray}
Here $N$ is the lattice size, $p_i = \frac{2\pi}{L} k_i, k_i = 0, ..., L-1$.

\begin{figure}
\begin{center}
\epsfig{figure=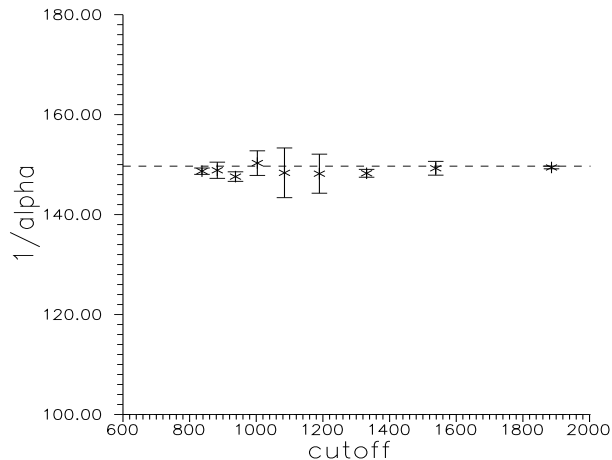,height=60mm,width=80mm,angle=0}
\caption{\label{alphel} The inverse renormalized fine structure constant
$1/\alpha(M_Z/\Lambda)$ as a function of the cutoff  $\Lambda$ at $\lambda
=0.0025$ , $\beta = 12$ on the lattice $16^3\times 32$. The dashed line is the
one - loop estimate for $\Lambda = 1$ TeV. }
\end{center}
\end{figure}

We  substitute to (\ref{V2}) the fit to the $Z$ - boson mass represented in
(\ref{fitmz}). The results are presented in Fig. \ref{alphel} and are to be
compared with the tree level estimate for the fine structure constant
$\alpha^{(0)} \sim \frac{1}{151}$ and the $1$ - loop approximation (when we
assume bare value of $\alpha$ to live at the scale $\sim 1$ TeV while the
renormalized value lives at the Electroweak scale $M_Z$): $\alpha^{(1)}(M_Z/{1
\, {\rm TeV}}) \sim \frac{1}{149.7}$.

The values of the renormalized fine structure constant calculated on the
lattice $16^3\times 32$ are close to the values calculated on the lattice
$8^3\times 16$  represented in \cite{PZ2011}. We observe that the renormalized
fine structure constant calculated in the mentioned above way is rather close
to the one - loop estimate (when the cutoff $\Lambda$ in
$\alpha^{(1)}(M_Z/\Lambda)$ is around $1$ TeV. It is worth mentioning, that the
present data on the renormalized fine structure constant (and the data of
\cite{PZ2011}) differ from that of reported in \cite{Z2010}. In \cite{Z2010} we
used the potential extracted from the Wilson loops and approximated it by the
simple $1/R$ fit. Moreover, in \cite{Z2010} the exchange by virtual Z - bosons
was neglected. Therefore the values represented in \cite{Z2010} depend strongly
on the lattice size and deviate essentially from the one - loop estimate near
to the phase transition point.

\section{Scalar field condensate}

\begin{figure}
\begin{center}
 \epsfig{figure=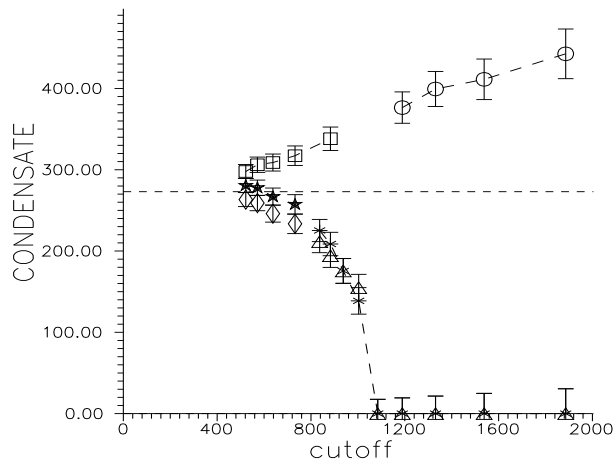,height=60mm,width=80mm,angle=0}
\caption{\label{fig.1_} The scalar field condensate (in GeV) as a function of
the cutoff
 for $\lambda =0.0025$ , $\beta = 12$.  Circles correspond to the UZ potential, lattice $16^3\times 32$.
Empty squares correspond to the UZ potential, lattice $8^3\times 16$. Crosses
correspond to the UDZ potential, lattice $16^3\times 32$. Stars correspond to
the UDZ potential, lattice $8^3\times 16$. Triangles correspond to the
ultraviolet potential, lattice $16^3\times 32$. Diamonds correspond to the
ultraviolet potential, lattice $8^3\times 16$.  }
\end{center}
\end{figure}

In \cite{PZ2011} three different effective constraint potentials were
introduced. All them are defined in Unitary gauge $H =
\left(\begin{array}{c}h\\0\end{array}\right)$ with real $h$. The first one is
the  ultraviolet potential
\begin{equation}
exp(-V^{u-v}(\phi)) = \langle \delta(\phi - h_x)\rangle\label{uv}
\end{equation}
Here real scalar field $h_x$ is defined on the lattice points $x$.

In order to define the infrared potential
\begin{equation}
exp(-V^{i-r}(\phi) )= \langle\delta(\phi - \frac{1}{N_4} \sum_x h_x)\rangle,
\label{ir}
\end{equation}
(where $N_4$ is the number of lattice points) it is necessary to  fix the
ambiguity
\begin{equation}
h_x \rightarrow (-1)^{n_x} h_x, \, Z \rightarrow [Z + \pi d n]{\rm mod}\, 2
\pi,
\end{equation}
where the Z - boson field is defined in (\ref{Z0_}).

The first way is minimization of
\begin{equation}
\sum_{links}(1-{\rm cos}\, Z) \rightarrow min \label{Z_}
\end{equation}
 with respect to the mentioned $Z_2$
transformations. In \cite{PZ2011} this gauge was called the $Z$ - version of
Unitary gauge and the corresponding effective potential (\ref{ir}) is called UZ
potential.

The second way to define the Unitary gauge with $h_x \in R$ is to minimize the
divergence of $Z$ with respect to the remaining $Z_2$ transformations:
\begin{equation}
\sum_x [\delta Z]^2 \rightarrow min \label{dz}
\end{equation}
This gauge is called the DZ - version of Unitary gauge and the corresponding
effective potential (\ref{ir}) is called UDZ potential.

The three mentioned above effective potentials give three different definitions
of the scalar filed condensate.  (The condensate $v$  is defined as the value
of $\phi$, at which the potential $V(\phi)$ has its minimum).

\begin{figure}
\begin{center}
 \epsfig{figure=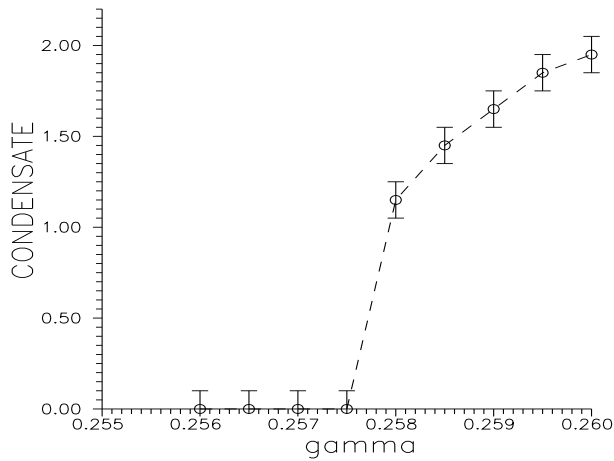,height=60mm,width=80mm,angle=0}
\caption{\label{qhim} The scalar field condensate extracted from the UZ
potential in lattice units as a function of $\gamma$
 for $\lambda =0.0025$ , $\beta = 12$ on the lattice $16^3\times 32$. }
\end{center}
\end{figure}

In Fig. \ref{fig.1_} we represent these three condensates as functions of the
cutoff $\Lambda = \frac{\pi}{a}$. We consider the condensates in physical
units, i.e. we multiply the values expressed in lattice units by
$\frac{\sqrt{\gamma}}{a}$. So, we define $v^{phys} = \frac{\sqrt{\gamma}}{a}
v$.

The scalar field condensate (in physical units) has to be renormalized. The
renormalized condensate is usually defined as $v^{phys}_R = \frac{1}{Z_H^{1/2}}
v^{phys}$, where the wave function renormalization constant enters the
following approximation for the scalar field propagator:
\begin{equation}
\langle [H_{p}]^{+} H_{q} \rangle - |\langle  H \rangle|^2  =
Z_{H}\frac{\delta_{pq} }{N_4(4 \sum_i {\rm sin}^2 p/2 + 4 \, {\rm sh}
^2\frac{m_H}{2})},\label{corMH_}
\end{equation}
Here  $H_p = \frac{1}{N_4} \sum_x e^{i px} H_x$, and $N_4 = N^3 L$.

Due to the renormalizability the Z - boson mass is related to the scalar field
condensate as follows: $M_Z =
 g_Z v^{phys}_R/2 = \frac{1}{Z^{1/2}_H} g_Z v^{phys}/2$, where $g_Z =
\frac{4 \sqrt{ \pi \alpha}}{{\rm sin} 2 \theta_W}$ is the renormalized coupling
constant. In the perturbation theory the deviation of $Z_H$ from unity is
proportional to the factor $\sim \alpha \, {\rm log} \frac{\Lambda}{M_Z} \sim
0.02$ (for $\Lambda \sim 1$ TeV). Therefore, for $\Lambda \sim 1 $ TeV we
expect $Z_H \sim 1$.

That's why the perturbation theory prompts that both the (nonrenormalized)
scalar field condensate (represented in Fig. \ref{fig.1_}) and the renormalized
one must be close to the value $v_0^{phys} = 2M_Z/g_Z = {\rm sin} 2 \theta_W
\,M_Z/\sqrt{4\pi \alpha} \sim 273$ GeV (this value differs from the
conventional value $246$ GeV due to the difference in $\alpha$). We observe
that all considered condensates approach this value when the cutoff is
decreased. However, an essential deviation from this value appears at the
values of the cutoff $\sim 1$ TeV. Looking at this plot we also conclude that
the condensates extracted from the UDZ potential and from the ultraviolet
potential represent close quantities\footnote{For $\Lambda < 0.8$ TeV we do not
have data from the lattice $16^3\times 32$. Therefore for these values of the
cutoff we represent on the plot the data obtained on the lattice $8^3\times
16$.}. At the same time the UZ potential gives different value of the scalar
field condensate. At the present moment we do not understand what is the reason
for such a behavior. It is worth mentioning that the condensate extracted from
the UZ potential vanishes at $\gamma_c^{\prime}$ (see Fig.\ref{qhim}). We must
remember, that the cutoff was not calculated at this point but Fig. \ref{chi}
indicates that at this point we may have $\Lambda_c^{\prime} \sim \infty$. (Let
us remind also that Fig.\ref{LAQ} points out to $\gamma_c^{\prime}$ as to the
phase transition point.) The other two condensates vanish at $\Lambda_c \sim 1$
TeV i.e. at $\gamma_c = 0.26075 \pm 0.00005$, where the value of the cutoff has
been calculated explicitly.  We shall see in the next section that close to
this point of the phase diagram both  the Z - string and the Nambu monopoles
begin to percolate.

The observed behavior of the scalar field condensates calculated in UZ and UDZ
- versions of Unitary gauge means that the wave function renormalization
constant for the scalar field (defined in these gauges) differs from the
perturbation theory prediction at $\Lambda > 800 $ GeV. (We may calculate this
constant as $Z_H = (\frac{273 \, {\rm GeV}}{v^{phys}})^2$, where $v^{phys}$ is
drawn in Fig.\ref{fig.1_}.)

\section{Z - strings and Nambu monopoles}

\begin{figure}
\begin{center}
 \epsfig{figure=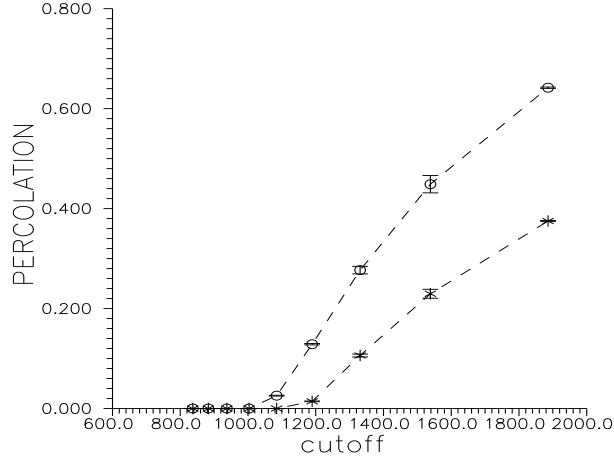,height=60mm,width=80mm,angle=0}
\caption{\label{fig.4} The percolation probability for the Z - string (circles)
and for the Nambu monopoles (crosses) as a function of the cutoff $\Lambda =
\frac{\pi}{a}$ (in GeV) at $\lambda =0.0025$ , $\beta = 12$ on the lattice
$16^3\times 32$.}
\end{center}
\end{figure}

In this section we use definition (\ref{Z1_}) of the Z - boson field. The
classical solution corresponding to a $Z$-string should be formed around the
$2$-dimensional topological defect which is represented by the integer-valued
field defined on the dual lattice $ \Sigma = \frac{1}{2\pi}^*([d Z_{{\rm mod}
2\pi} - d Z)$. Therefore, $\Sigma$ can be treated as the worldsheet of a {\it
quantum} $Z$-string \cite{Chernodub_Nambu}. Then, the worldlines of quantum
Nambu monopoles appear as the boundary of the $Z$-string worldsheet: $ j_Z =
\delta \Sigma $.

The percolation probability of Nambu monopoles is defined as follows. First let
us denote the probability that two points $x$, $y$ are connected by the
monopole cluster by $\rho(x,y)$. We may identify it with the following quantity
\begin{equation}
\langle \Psi^+_x \Psi_y \rangle = \rho(x,y),\label{psi}
\end{equation}
where operator $\Psi^+_x$ creates the monopole - antimonopole pair at the point
$x$. This identification allows us to calculate the lightest monopolium mass
$m_M$, i.e. the mass of the quantum state consisted of the monopole -
antimonopole pair connected by the Z - string:
\begin{eqnarray}
\Pi_1(|x_0-y_0|) & = & \frac{1}{N^6} \sum_{x_1,x_2,x_3,y_1,y_2,y_3}
\langle \Psi^{+}_{x} \Psi_{y} \rangle  \nonumber\\
&\sim &
  A\, (e^{-m_{M}|x_0-y_0|}+ e^{-m_{M}(L - |x_0-y_0|)})
\label{corM}
\end{eqnarray}
Here the lattice size is $N^3\times L$, and it is implied that the mass is
calculated in the region of the phase diagram, where the condensate of $\Psi$
vanishes. In order to calculate this condensate we consider the following
quantity:
\begin{eqnarray}
\Pi_3(|\bar{x}-\bar{y}|) & = & \frac{1}{N^2} \sum_{x_3,y_3}
\langle \Psi^{+}_{(\bar{x}, x_3)} \Psi_{(\bar{y},y_3)} \rangle  \nonumber\\
&\sim & |\langle \Psi \rangle|^2 + \tilde{\Pi}_3(|\bar{x}-\bar{y}|),
\label{corM_}
\end{eqnarray}
where $ \bar{x} = (x_0,x_1,x_2)$, and  $ \tilde{\Pi}_3(r) \rightarrow 0 \, (r
\rightarrow \infty)$. Thus the condensate is defined through the percolation
probability: ${C}_{\mathrm{mon}} =|\langle \Psi \rangle|^2 = {\rm lim}_{r
\rightarrow \infty} \Pi_3(r)$. It is worth mentioning that the two quantities
$\Pi_1 (r)$ and $\Pi_3 (r)$  give different values at $r \rightarrow \infty$ if
there is the massless scalar excitation in the spectrum, otherwise $\Pi_1 (r) -
\Pi_3 (r) \rightarrow 0 \, (r \rightarrow \infty)$.

In a similar way we are also able to calculate the mass of the lightest
excitation
 of the Z - string. First, the probability of the two points to be connected by
 the Z - string cluster is defined. Next, this probability is related to the
two - point correlator of the operators that create the Z - string excitations.
After that the mass of the lightest excitation is extracted and the percolation
probability for the Z - strings is  defined. The percolation probabilities for
the Nambu monopoles and the Z - strings are represented in Fig. \ref{fig.4}. We
observe that both monopole currents and Z - string worldsheets percolate at
$\Lambda > 1.1$ TeV.

\begin{figure}
\begin{center}
 \epsfig{figure=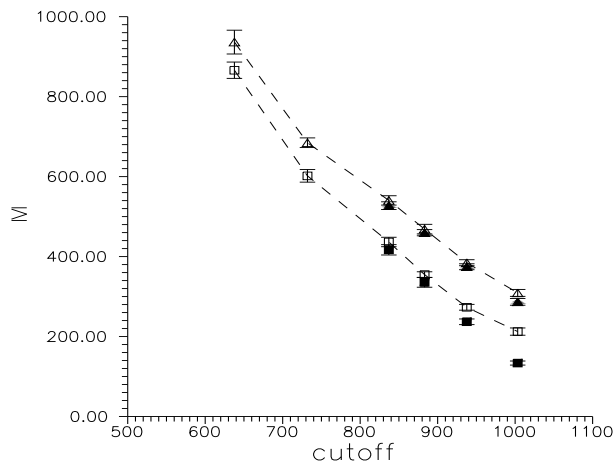,height=60mm,width=80mm,angle=0}
\caption{\label{massms} The monopolium mass (tiangles) and the lightest string
excitation mass (squares) as a function of the cutoff calculated at $\lambda
=0.0025$ , $\beta = 12$ on the lattice $8^3\times 16$ (empty symbols) and
$16^3\times 32$ (dark symbols).}
\end{center}
\end{figure}

The monopolium mass and the lightest Z-string excitation mass calculated  are
represented in Fig. \ref{massms} as  functions of the cutoff. One can see, that
both these masses decrease when the cutoff is increased. At  $\Lambda > 1.1$
TeV we do not calculate the mentioned masses because the condensates of the
monopolium and of the Z string excitations appear.

The code for the calculation of the percolation probability was written
especially for the investigation reported in this paper. It has been tested in
several ways. In particular, results of \cite{Gubarev} and \cite{BVZ2006} were
reproduced.

\begin{figure}
\begin{center}
 \epsfig{figure=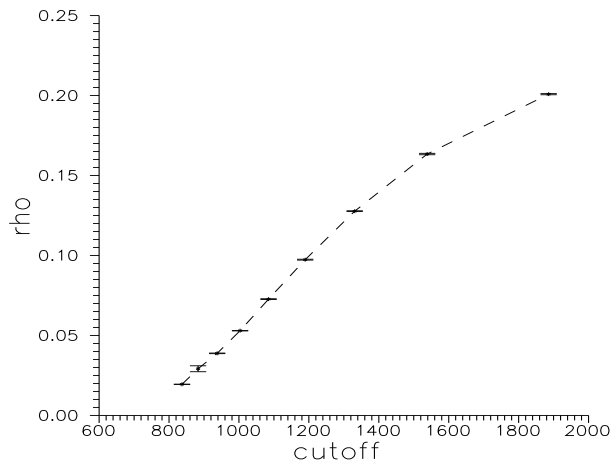,height=60mm,width=80mm,angle=0}
\caption{\label{fig.5} The Nambu monopole density (in lattice units) as a
function of the cutoff  at $\lambda =0.0025$ , $\beta = 12$ on the lattice
$16^3\times 32$.}
\end{center}
\end{figure}

According to the classical picture the Nambu monopole size is of the order of
$M^{-1}_H$. Therefore, for example, for $a^{-1} \sim 250$ GeV and $M_H \sim
 150$ GeV the expected size of the monopole is about two lattice
spacings. The monopole density around $0.015$ means that among about $16$ sites
there exists $1$ site that is occupied by the monopole. Average distance
between the two monopoles is, therefore, about $2$ lattice spacings that is the
monopole size. In Fig. \ref{fig.5} (where the Nambu monopole density is
represented as a function of the cutoff) we observe, that at $\Lambda
> \Lambda_{c2} \sim 0.8$ TeV the Nambu monopole density is larger than $0.015$,
i.e. the average distance between monopoles is less then the  classical
monopole size. According to \cite{Z2010} this means that at $\Lambda_{c2}$ we
enter the fluctuational region. It is worth mentioning that within this region
the notion of quantum Nambu monopole differs from the notion of the classical
Nambu monopole considered in \cite{Nambu}. In particular, the size of the
quantum object may be sufficiently less than that of the classical Nambu
monopole.

The original estimate of the Nambu monopole mass was given by Nambu
\cite{Nambu}:
\begin{equation}
M_N \sim \frac{4\pi}{3e}\, {\rm sin}^{\frac{5}{2}} \, \theta_W \,
\sqrt{\frac{M_H}{M_W}} \, 246\, {\rm GeV} \sim 900 \, {\rm GeV} \label{MN}
\end{equation}

Then, according to \cite{Nambu} the classical energy of  monopole -
antimonopole pair is
\begin{equation}
 E = 2M_M -\frac{Q^2}{4\pi l},\label{Mm_}
\end{equation}
where $Q = \frac{4\pi}{e\, {\rm sin}^2 \theta_W}$ is the monopole charge while
$l$ is the average distance between the monopoles in the monopolium. We can use
this formula to estimate roughly the dependence of the monopolium mass on the
cutoff. At small values of $\Lambda$ the contribution of the $1/l$ term can be
neglected (as the monopole density is negligible) and the lightest monopolium
mass is expected to be close to the value $\sim 1.8$ TeV. However, when the
cutoff is increased, the average distance between Nambu monopoles is decreased.
Therefore, $l$ in (\ref{Mm_}) is decreased as well. As a result the monopolium
mass is decreased. Indeed, we observe this kind of behavior in Fig.
\ref{massms}.

It is worth mentioning that the monopolium and the Nambu monopole itself are
unstable as classical objects \cite{Nambu}. However, when the cutoff is
increased, the whole picture of Nambu monopoles and monopolium is changed. The
operator (defined in (\ref{psi})) that creates the monopole - antimonopole pair
creates actually the quasiparticle, the properties of which may differ
essentially from the properties of the classical monopolium. In particular,
these quasiparticles are condensed at $\Lambda > 1.1$ TeV.

\section{Discussion}

\label{disc}

 Let us try to estimate the conditions under which the
perturbation theory can be applied to this model. We make this estimate in the
spirit of volume 5, paragraph 146 of \cite{Landau} (where the similar
considerations were used in order to estimate the width of the fluctuational
region in the finite temperature Ginzburg - Landau model). We are going to
compare the vacuum average of $H$ with the fluctuations of $H$ within the 4-
volume $N_4$, the linear size of which is equal to the correlation length of
$H$. This correlation length in lattice units is equal to $1/m_H$. The
fluctuations are obtained from (\ref{IRPOT_}) and are of the order of $\delta H
\sim \frac{1}{v\sqrt{8N_4\lambda}}\sim \frac{m_H^2}{v\sqrt{8\lambda}}$.
Therefore, at $v >> \frac{m_H^2}{v\sqrt{8\lambda}}$ the perturbation theory can
be applied while otherwise it might not be applied. Finally, we obtain $\lambda
<< \gamma^2/8$ or $\frac{m_H}{m_W}\sqrt{4\pi \alpha} << 1$. For $m_H \sim 150$
GeV and $\alpha \sim 1/150$ this expression reads $0.54 << 1$. This estimate is
indeed confirmed by numerical results (see the end of  Section 2). Thus already
on this level there may appear some doubts about the validity of the
perturbation expansion within the given model.

Above we have reported the results of our numerical investigation of lattice
Weinberg - Salam model at $\beta = 12$, $\lambda = 0.0025$, $\theta_W = 30^o$.
For these values of couplings the bare Higgs boson mass is close to $150$ GeV
near to the transition between the Higgs phase and the symmetric phase.
Numerical simulations were performed on the lattices of sizes up to $16^3\times
32$.

Our data draw the following pattern of the phase transition.

1. When the cutoff is increased ($\gamma$ is decreased) $Z$ vortices become
more and more dense. Somewhere around $\Lambda \sim 0.8$ TeV ($\gamma_{c2}\sim
0.2625$) the transition to the fluctuational region occurs \cite{Z2010}. In
this region $Z$ - vortices and the Nambu monopoles dominate. The average
distance between Nambu monopoles becomes compared to their sizes.

2. At the value of $\Lambda$ around $\Lambda_c \sim 1$ TeV  ($\gamma$ around
$\gamma_c \sim 0.26075$) the scalar field condensates calculated using the UDZ
effective potential and the ultraviolet effective potential vanish. At $\Lambda
> 1.1 $ Tev the Nambu monopoles and the
Z - strings begin to percolate. This means, in particular, that the operator
that (naively) creates the so - called monopolium state actually creates the
quasiparticles that are condensed.

3. At the value of $\gamma$ around $\gamma_c^{\prime} \sim 0.25775$ the scalar
field condensate calculated using the UZ effective potential vanishes. Also
somewhere close to $\gamma_c^{\prime}$ the derivative of the link part of the
action has the step - like discontinuity. We cannot calculate the ultraviolet
cutoff at $\gamma_c^{\prime}$ due to large statistical errors. At the present
moment we do not exclude that it tends to infinity at this point on the ideal
infinite lattice.

The technical question about the order of the phase transition remains. There
still exist two possibilities: either we deal with the second order phase
transition (localized at $\gamma_c^{\prime}$) or with the crossover. The first
possibility is realized if at $\gamma_c^{\prime}$ all lattice masses vanish or
the correlation lengths are infinite at this point. The behavior of the scalar
field fluctuation indicates that this may indeed be true. However, accurate
investigation of the lattice masses is still to be performed in the vicinity of
$\gamma^{\prime}_c$. Let us now estimate the lattice size needed for such an
investigation. Suppose, we need to investigate the region of the phase diagram
with $\Lambda = \frac{\pi}{a} \sim 1$ TeV. Then, the lattice size has to be
much larger than the correlation length: $L >> \frac{1}{m_H} =
\frac{1}{150\,{\rm GeV}\, a} \sim 2$. Therefore, the lattices of sizes
$16^3\times 32$ seem to us large enough to investigate the model at $\gamma \ge
0.2585$.  If, however, we are going to investigate the region of the phase
diagram with $\Lambda \sim 10$ TeV, we need to have lattices of sizes $L
>> 10$. For this purpose lattices used in the present research are not large
enough.   That's why if the second order phase transition is indeed present at
$\gamma_c^{\prime}$, in a small vicinity of this point, where $\Lambda >> 1$
TeV (most likely, this vicinity is situated within the interval $[0.2575,
0.2585)$), the numerical lattice methods that use lattices of sizes up to
$16^3\times 32$ cannot be applied for the calculation of lattice masses.

The scalar field condensates calculated in three different ways in our study
differ from each other and from the expected value $\sim 273$ GeV for the
values of $\Lambda > 800 $ GeV. However, all them have tendencies to approach
this value when $\Lambda$ is decreased. This means that the wave function
renormalization constant for the scalar field differs from its perturbative
estimate  at $\Lambda > 800 $ GeV. The percolation probability for the Nambu
monopoles and for the $Z$ - strings differ from zero at $\Lambda
> 1 $ TeV. We consider this behavior as a manifestation of the nonperturbative
effects present in the given model at large enough energy scales. It is worth
mentioning that the point of view that the nonperturbative effects may become
important in the Higgs sector of the Standard Model is not new. In particular,
in \cite{Consoli} it was argued that the wave function renormalization constant
for the scalar field contains large nonperturbative contribution (at zero
temperature). This conclusion of \cite{Consoli} is in accordance with our
results represented here.

The situation seems to us similar to the phase transition in the second order
superconductors at finite temperature. Namely, there exists the fluctuational
region around the critical temperature $(T_c - \Delta T; T_c + \Delta T)$,
where the perturbation theory cannot be applied and the nonperturbative effects
are present \cite{Landau}. In the lattice Weinberg - Salam model (at zero
temperature) there also exists such a region around the phase transition at
$\gamma_c^{\prime}$. This region is localized within the interval
$[\gamma_c^{\prime}; \gamma_{c2})$ that corresponds to values of the cutoff
$\Lambda > 0.8$ TeV. As it was explained above we observe indications that
within this region nonperturbative effects become important.

\begin{acknowledgments}

The author kindly acknowledges discussions with M.I.Polikarpov, P.Buividovich
and E. Luschevskaya. This work was partly supported by RFBR grant 09-02-00338,
11-02-01227, by Grant for Leading Scientific Schools 679.2008.2. This work was
also supported by the Federal Special-Purpose Programme 'Cadres' of the Russian
Ministry of Science and Education. The numerical simulations have been
performed using the facilities of Moscow Joint Supercomputer Center,
 the supercomputer center of Moscow University, and the supercomputer center of Kurchatov
 Institute.

\end{acknowledgments}

\end{document}